\documentclass[sigconf]{acmart}

\AtBeginDocument{%
  \providecommand\BibTeX{{%
    \normalfont B\kern-0.5em{\scshape i\kern-0.25em b}\kern-0.8em\TeX}}}

\acmYear{2022} 
\setcopyright{rightsretained} 
\acmConference[MM '22]{Proceedings of the 30th ACM International Conference on Multimedia}{October 10--14, 2022}{Lisboa, Portugal}
\acmBooktitle{Proceedings of the 30th ACM International Conference on Multimedia (MM '22), October 10--14, 2022, Lisboa, Portugal}
\acmDOI{10.1145/3503161.3548411}
\acmISBN{978-1-4503-9203-7/22/10}

\makeatletter
\newcommand\footnoteref[1]{\protected@xdef\@thefnmark{\ref{#1}}\@footnotemark}
\makeatother
\usepackage{multirow}

\copyrightyear{2022} 
\acmYear{2022} 
\setcopyright{rightsretained} 
\acmConference[MM '22]{Proceedings of the 30th ACM International Conference on Multimedia}{October 10--14, 2022}{Lisboa, Portugal}
\acmBooktitle{Proceedings of the 30th ACM International Conference on Multimedia (MM '22), October 10--14, 2022, Lisboa, Portugal}
\acmDOI{10.1145/3503161.3548411}
\acmISBN{978-1-4503-9203-7/22/10}
\usepackage{etoolbox}
\makeatletter
\patchcmd{\maketitle}{\@copyrightpermission}{
   \begin{minipage}{0.3\columnwidth}
     \href{https://creativecommons.org/licenses/by/4.0/}{\includegraphics[width=0.90\textwidth]{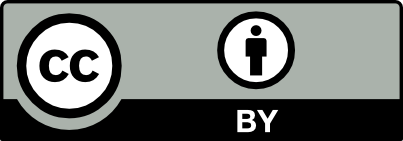}}
   \end{minipage}\hfill
   \begin{minipage}{0.7\columnwidth}
     \href{https://creativecommons.org/licenses/by/4.0/}{This work is licensed under a Creative Commons Attribution International 4.0 License.}
   \end{minipage}

   \vspace{5pt}
}{}{}

\makeatother

\begin{document}

\title{MM-ALT: A Multimodal Automatic Lyric Transcription System}

\author{Xiangming Gu}
\authornote{Both authors contributed equally to this research.}
\affiliation{%
  \institution{Integrative Sciences and Engineering Programme, NUS Graduate School, National University of Singapore}
  \country{Singapore}
  }
\email{xiangming@comp.nus.edu.sg}

\author{Longshen Ou}
\authornotemark[1]
\affiliation{%
  \institution{School of Computing, National University of Singapore}
  \country{Singapore}}
\email{longshen@comp.nus.edu.sg}

\author{Danielle Ong}
\affiliation{%
  \institution{School of Computing, National University of Singapore}
  \country{Singapore}
  }
\email{danielleong@u.nus.edu}

\author{Ye Wang}
\affiliation{%
  \institution{School of Computing, National University of Singapore}
  \country{Singapore}
}
\email{wangye@comp.nus.edu.sg}




\begin{abstract}

Automatic lyric transcription (ALT) is a nascent field of study attracting increasing interest from both the speech and music information retrieval communities, given its significant application potential. However, ALT with audio data alone is a notoriously difficult task due to instrumental accompaniment and musical constraints resulting in degradation of both the phonetic cues and the intelligibility of sung lyrics. To tackle this challenge, we propose the MultiModal Automatic Lyric Transcription system (MM-ALT), together with a new dataset, N20EM, which consists of audio recordings, videos of lip movements, and inertial measurement unit (IMU) data of an earbud worn by the performing singer. We first adapt the wav2vec 2.0 framework from automatic speech recognition (ASR) to the ALT task. We then propose a video-based ALT method and an IMU-based voice activity detection (VAD) method. In addition, we put forward the Residual Cross Attention (RCA) mechanism to fuse data from the three modalities (i.e., audio, video, and IMU). Experiments show the effectiveness of our proposed MM-ALT system, especially in terms of noise robustness. Project page is at \url{https://n20em.github.io}.

\end{abstract}

\ccsdesc[500]{Applied computing~Sound and music computing}
\ccsdesc[500]{Information systems~Music retrieval}
\ccsdesc[500]{Information systems~Speech / audio search}
\ccsdesc[300]{Computing methodologies~Neural networks}
\ccsdesc[100]{Human-centered computing~Ubiquitous and mobile computing systems and tools}

\keywords{lyric transcription, multimodality, neural networks, dataset}

\maketitle

\section{Introduction}

Automatic lyric transcription (ALT) is the task of recognizing lyric text from singing. The supplementing of textual information via ALT often facilitates solutions to other music information retrieval problems, such as lyric alignment \cite{gupta2020alt:bgm}, query by singing \cite{hosoya2005querybysing}, audio indexing \cite{fujihara2008audioindexing}, and music subtitling \cite{dzhambazov2017subtitle}. Additionally, ALT systems can be augmented to appraise singing from various linguistic or musical aspects, such as word intelligibility and pronunciation\cite{murad2018slions}. Recently, research in ALT has become increasingly active, as evidenced by the emergence of benchmark singing datasets, e.g., \cite{dabike2019alt:dsing, meseguer2019dali1, meseguer2020alt:dali2}, and further studies of advanced acoustic modeling techniques, e.g., \cite{basak2021alt:e2e, ahlback2021alt:mstre}. 


To our knowledge, ALT systems thus far are built with audio-only data, which has arguably led to a plateau in research efforts plagued by the following reoccurring challenges:

\textbf{Difficulty of the ALT task}.
Since musicality distinguishes singing from speech, a singer inevitably sacrifices some richness of linguistic features such as word stress and articulation to compensate for musical features such as melody, tempo, and timbre. Consequently, singing is generally less intelligible than speech \cite{sharma2019sing_eval}. This insufficient richness of linguistic features handicaps approaches adapted from automatic speech recognition (ASR) to ALT. This might account for the performance gap (19.60\% vs. 8.7\% word error rate) of the same acoustic model between ALT \cite{dabike2019alt:dsing} and a similar-scale ASR task \cite{povey2018asr:tdnnf}.

\textbf{Nonconformity with human perception patterns}.
The perception of speech extends beyond the auditory realm, as observed by the McGurk effect \cite{mcgurk1976} that visual information can substantially contribute to auditory perception. Studies have shown that visual cues of speech play an essential role in language learning \cite{meltzoff1977imitation, davies2009investigating}, lending further support to the notion that human auditory perception relies on multimodal information. 

\textbf{Insufficient robustness}.
Given the two points above, it is reasonable to surmise that attempting ALT on audio-only recordings in challenging signal-to-noise ratio (SNR) environments would probably yield unsatisfactory results. Coupled with lowered intelligibility over speech and the lack of external cues, the added factor of noise (e.g., accompaniments) further increases the difficulty of retrieving information from musical audio. 

Due to the interconnected relationship between sound production and observable lip movements, it is clear that adding the modality of visual information to the ALT system is a promising prospect at present. Recent advances in audio-visual speech recognition (AVSR), e.g., \cite{ma2021avsr:conformer, shi2022avsr:robust}, show that audio-visual models perform better than their audio-only counterparts, suggesting that including additional sensory information on top of the acoustic signal enhances the overall performance of ALT systems by enriching the input to the model. We assume that training an ALT system on multimodal data is a viable solution to the challenges mentioned earlier. 

Additionally, inertial measurement unit (IMU) data from wearable devices are an augmentation offering potential for improving system performance. In the early stages of data collection, we observed that participants tended to make rhythmic head movements and facial expressions along with the music, in addition to regular pronunciation efforts during singing. This was especially obvious for those who displayed higher levels of musicality. Furthermore, the jaw movements of singers are highly correlated with the pronunciation of lyrics. We assume that these additional cues are valuable in improving lyric recognition and increasing system robustness. We use \textit{eSense} \cite{Kawsar2018esense}, an earbud augmented with inertial sensors, to capture data of facial and head movements. This device has been proven to reliably detect head-, jaw-, and mouth-related movements \cite{Purabi2019head, Rupavatharam2019jaw, Lotfi2020chew}, facial expression \cite{Lee2019facial}, speakers in conversations \cite{Min2018speaker}, and even human emotions \cite{Purabi2019head}.

We conduct this work to validate our assumptions. The contributions are summarized into three aspects:
\begin{itemize}
\item We present the MultiModal Automatic Lyric Transcription system (MM-ALT), which utilizes three modalities of input: audio, video, and signals from wearable IMU sensors. To facilitate building the system, we curate the N20EM dataset for multimodal lyric transcription. We create a group of models on this dataset that can perform multimodal lyric transcription with varying combinations of modalities, obtaining a minimum word error rate (WER) of 12.71\%. We further reveal an increase in system robustness by introducing additional modalities. With severe perturbations of musical accompaniments (-10 dB SNR), our system can achieve 27.04\% absolute lower WER compared to its audio-only counterparts. 

\item We initialize two new tasks: lyric lipreading and IMU-based voice activity detection (VAD). In the lyric lipreading task, we attempt to recognize lyrics in singing utilizing only video information. Our video encoder is the first attempt to retrieve language-related information from singing recordings without the help of audio input. As to the IMU-VAD task, our IMU encoder is the first attempt at building a frame-level VAD system solely from motion data captured by a wearable IMU device. Our experiments elucidate the correlation between the IMU and audio modalities. 



\item We propose Residual Cross Attention (RCA), a new feature fusion method to better fuse the multimodal features using self-attention and cross-attention mechanisms. We demonstrate the effectiveness of this new modality fusion method in our ALT system by comparing it with various feature fusion methods.
\end{itemize}
\section{Related Work}

\subsection{Automatic Speech Recognition}

The development of deep neural network (DNN) techniques has revolutionized ASR by prompting a shift in paradigm from hidden-Markov-model- (HMM-) based systems, to DNN-HMM hybrid models, and finally, to end-to-end (E2E) models \cite{li2022RecentAdvancesEndtoEnd}. Compared to previous models, E2E ASR systems possess more consistent optimization objectives, more simplified pipelines, and more compact model structures\cite{li2022RecentAdvancesEndtoEnd}. The Connectionist Temporal Classification (CTC) \cite{graves2006e2e:ctc} and Attention-based Encoder-Decoder (AED) \cite{cho2014e2e:aed1, bahdanau2014e2e:aed2} are two mainstream E2E ASR models. Jointly optimizing the shared encoder of AED and CTC models in a multitask learning framework greatly improves convergence and mitigates the misalignment issue of AED models \cite{kim2017asr:joint1, hori2017asr:joint2}. 

Training state-of-the-art ASR models under supervised learning framework requires a large amount of transcribed data, which is too demanding for low-resource languages. In recent years, it has become a trend to use self-supervised learning (SSL) techniques to build ASR systems. By utilizing easy-to-access unlabeled data, SSL models can learn general data representations to facilitate downstream tasks. Models such as wav2vec 2.0 \cite{baevski2020wav2vec}, autoregressive predictive coding \cite{chung2020ssl:apc}, and HuBERT \cite{hsu2021hubert} can learn very powerful semantic representations of audio. These models can achieve impressive results even with very limited labeled data, demonstrating the feasibility of low-resource speech recognition.

It is convenient to incorporate multimodal input into E2E ASR models, especially a simultaneous audio and video signal. Audio-visual speech recognition (AVSR) brings machines closer to how humans perceive speech. Lots of research supports the notion that adding video input to the ASR systems has positive effects on recognition performance, e.g., \cite{petridis2018avsr:e2e, zhou2019avsr:mmattn, tao2020avsr:mtl, ma2021avsr:conformer, shi2022avsr:robust}. In particular, Audio-Visual Hidden Unit BERT (AV-HuBERT) \cite{shi2022avsr:avhubert, shi2022avsr:robust} notably outperforms previous state-of-the-art models in both lipreading and AVSR tasks by learning the joint representation of a synchronized speech and audio signal.

\subsection{Automatic Lyric Transcription}
ALT is the counterpart problem of ASR in the field of music information retrieval. The research of ALT started from \cite{hosoya2005querybysing}, by adapting an HMM model to build a Japanese lyric transcription system. Then, Mesaros and Virtanen studied the influence of in-domain lyric language models (LM) \cite{mesaros2010RecognitionPhonemesWords}. \cite{mcvicar2014alt:repetition} leveraged musical structure to solve the ALT problem by using the repetitive chorus of one song to improve the consistency and accuracy of their ALT system. Afterward, Kawai et al. first attempted a deep learning approach to ALT by a DNN-HMM model \cite{kawai2016SpeechAnalysisSungspeech}. In the baseline system constructed for the DSing dataset \cite{dabike2019alt:dsing}, a contemporary state-of-the-art acoustic model TDNN-F was used for ALT. Demirel et al. then utilized a convolutional neural network (CNN) and time-restricted self-attention to extract more robust features before feeding the input to the TDNN-F \cite{demirel2020AutomaticLyricsTranscription}. Gupta et al. tried building a genre-informed acoustic model for ALT systems directly on polyphonic singing audio \cite{gupta2020alt:bgm}. Demirel et al. also built a multi-stream TDNN-F model \cite{ahlback2021alt:mstre} that performed better than the original version. More recent network architectures such as Transformers have also been used in ALT systems \cite{basak2021alt:e2e}.



One reason for the slower progress in developing ALT systems compared to advancements in ASR, is the inaccessibility of large-scale singing datasets \cite{zhang2021alt:pdaugment}. Therefore, some researchers have focused on providing basic support for the ALT problem from a data standpoint. For example, Dabike et al. built a pipeline to preprocess the DAMP Sing! 300x30x2 dataset \cite{smulesing300} to form an utterance-level dataset suitable for ALT \cite{dabike2019alt:dsing}, and Meseguer-Brocal et al. built a large-scale popular music dataset, DALI, containing synchronized audio, lyrics, and notes \cite{meseguer2019dali1, meseguer2020alt:dali2}.

Another approach to alleviate the problem of insufficient data is data augmentation. For instance, Kruspe adjusted speech data to make it more "song-like" by random time stretching and pitch adjusting \cite{kruspe2015TrainingPhonemeModels}. Basak et al. used a vocoder-based synthesizer to generate singing voices according to natural speech \cite{basak2021alt:e2e}. Lastly, Zhang et al. proposed aligning lyrics with melodies before adjusting duration and pitch during data augmentation \cite{zhang2021alt:pdaugment}.

\subsection{Multimodal Learning}

Humans perceive the world via different modalities, \textit{e.g.}, vision, sound, touch, and smell. These modalities have heterogeneous and complementary information, enabling humans to understand their surroundings better. Therefore, it is natural to design multimodal systems to process and fuse multiple modalities' high-dimensional inputs simultaneously. Given recent advancements in deep learning techniques, multimodal methods have surpassed their single-modal counterparts in empirical applications. For example, \cite{hu2019acnet, seichter2021efficient} combined RGB and depth images to improve semantic segmentation; \cite{shi2022avsr:avhubert, ma2021avsr:conformer, tao2021someone, afouras2018avsr:deepe2e, shi2022avsr:robust} adopted video and audio modalities to boost the performances of ASR and active speaker detection systems; \cite{trumble2017total, huang2020deepfuse} fused multi-view images or videos and IMU data to benefit 3D human pose estimation. Apart from empirical results in previous literature, Huang et al. proceeded to formalize the multimodal learning problem into a theoretical framework. They proved that learning with multiple modalities results in more accurate estimated representations of latent spaces, which can achieve better performance of learning than using its subset of modalities \cite{huang2021makes}. 

One fundamental problem in multimodal learning is to integrate multimodal data in a way that can exploit the complementary relationship and redundancy of multiple modalities \cite{baltruvsaitis2018multimodal}. A common technique is to fuse individual modality inputs or intermediate features together. This fusion operation could be concatenation, e.g., \cite{wang2020makes, shi2022avsr:avhubert, ma2021avsr:conformer, trumble2017total}, addition, e.g., \cite{huang2020deepfuse, hu2019acnet, seichter2021efficient}, or  incorporation of an attention mechanism, e.g., \cite{tao2021someone, chen2021multimodal, yao2020multimodal, zhang2019neural, yang2020visual}. Given the success of transformers in computer vision \cite{dosovitskiy2020image, liu2021swin}, natural language processing \cite{vaswani2017attention, devlin2018bert} and speech processing \cite{baevski2020wav2vec, hsu2021hubert}, transformers are now widely used to learn multimodal representations. Among them, Perceiver models iteratively distill multimodal inputs into a tight latent bottleneck and build latent transformers to process these latent features \cite{jaegle2021perceiver1, jaegle2021perceiver2, carreira2022hierarchical}. Nagrani et al. proposed the framework of a Multimodal Bottleneck Transformer (MBT), which fuses the multimodal data via attention bottlenecks \cite{nagrani2021attention}. 

\section{Dataset}
This section describes the construction process and the preprocessing procedures of our N20EM dataset.

\subsection{Corpus Curation} 
We adopted the same 20 songs used in \cite{NUS48-Ecorpus} for practical reasons of rich phonemic coverage, ease of learning, and variation in musical features like genre and tempo. Instead of excising repeated lyrics and scat singing, we opted to preserve all utterances for completeness of data and naturalness of singing from participants. Seventeen males and thirteen females were recruited from a local university, with musical backgrounds ranging from no formal vocal training to amateur level exposure. Participants spoke English with accents including North American, European, Indian, East Asian, and Southeast Asian. Participants were free to choose songs they were more familiar and comfortable with singing from the 20 songs. However, we limited each song only to be selected by a maximum of 10 participants to encourage a greater selection diversity within each participant's repertoire and achieve a more balanced dataset. 

\subsection{Data Collection Procedure} 

All data were recorded in a soundproof studio. Our setup consisted of three main components: 

\begin{itemize}
\item An Audio-Technica 4050 condenser microphone with a pop filter was used to record audio data at 32-bit depth and 44.1 kHz sampling rate using Adobe Audition. A monaural headset was used for playback of musical accompaniment to participants during singing. 
\end{itemize}

\begin{itemize}
\item An earbud, \emph{eSense} \cite{Kawsar2018esense}, containing several built-in components, including an IMU sensor, a speaker, and a microphone, was used to record the kinesthetic activity of the head, jaw, and lips during recording at a sampling rate of 100 Hz. 
\end{itemize}

\begin{itemize}
\item A Sony AX4 video camera and a ring light were placed in front of participants, focused on the lower half of each singer's face, to capture footage of movements of oral articulators (jaw, lips, tongue) at 1920x1080 pixels, at 50 Hz frame rate. 
\end{itemize}

Participants were instructed to minimize bodily movements such as hand gestures and body swaying to reduce noise factors in data, such as accidental obstruction of video footage or anomalous variance in the sensory data. Lyric sheets for all songs were printed and placed on a music stand by the microphone for the subject's reference. The tempo and key of each song were predetermined. For comfort in vocal range, participants were given the option of a male-vocals arrangement or female-vocals arrangement for songs with greater variance in melodic range. Although the backing tracks fed to participants included vocals, participants were given some freedom in the rendition of pitch and rhythm during singing. When a track with all the lyrics clearly sung was obtained, the subject proceeded to the next song. A few pronunciation errors were allowed as long as the utterance remained clear.

Since singing is often studied in tandem with speech, we also collected read versions of the song lyrics for future comparative studies. For each song, the selected lyrics were read first, then sung with the aid of a backing track in a separate recording afterward. We believe this is beneficial for future research on the quantitative difference between singing and parallel speech\footnote{\label{note1}Presently, the parallel speech recordings are not used in the study since the focus of this paper is to build an ALT system.}.

\subsection{Data Preprocessing}

Data from different modalities were first synchronized with 40 ms precision. For video recordings, following the work of \cite{shi2022avsr:avhubert}, we first down-sampled videos to 25 Hz, then cut out 96 $\times$ 96 Regions-Of-Interest (ROIs) centered around the mouth. Only cropped videos containing lip areas will appear in the public dataset to protect participant privacy. Audio data were down-sampled to 16 kHz to match the input specification of wav2vec 2.0 \cite{baevski2020wav2vec}. 

We opted for an utterance-level annotation of lyrics. First, we manually annotated the starting and ending timestamps of each utterance. The standard of annotation was determined mainly by natural factors such as musical cadence or practical aspects such as preferring consonant boundaries over vowel boundaries between utterances. The annotation was done using spectrogram information and the marker function in Adobe Audition. Second, actually sung words were transcribed to serve as the ground truth for ALT. When they were different from the correct lyrics that should be sung, we used a set of notations to mark different types of errors. We believe this information will be helpful for future singing pronunciation evaluation research on this dataset\footnote{The error annotations are not utilized in the current system.}. Please refer to Appendix A for the annotation details.

After obtaining the annotation, the recordings of different modalities were all segmented to utterance level according to the annotation. If any, silence, breaths, or non-phonemic noise in between utterances were excised from the data. Then, the dataset was split into training, validation, and test sets. We ensure that no subset contained utterances from the same song to test the generalization abilities of the ALT systems. Table \ref{tb:split} shows the statistics of different subsets. 

\begin{table}[t]
\caption{Division of data into different sets for ALT system}
\centering
 \begin{tabular}{ l | c | c } 
  \toprule
  Set & Duration & Number of Utterances \\ 
  \midrule
  Total & 5 h 23 min & 5116 \\ 
  Train & 4 h 1 min & 3803 (74\%) \\ 
  Validation & 35 min & 616 (12\%) \\ 
  Test & 47 min & 697 (14\%) \\ 
  \bottomrule
 \end{tabular}
 
 \label{tb:split}
\end{table}

\section{MM-ALT System}
In this section, we first formalize the task of multimodal ALT and then describe our proposed MM-ALT for addressing the problem.
\begin{figure*}[t!]
\vskip -0.1in
\begin{center}
\includegraphics[width=\linewidth]{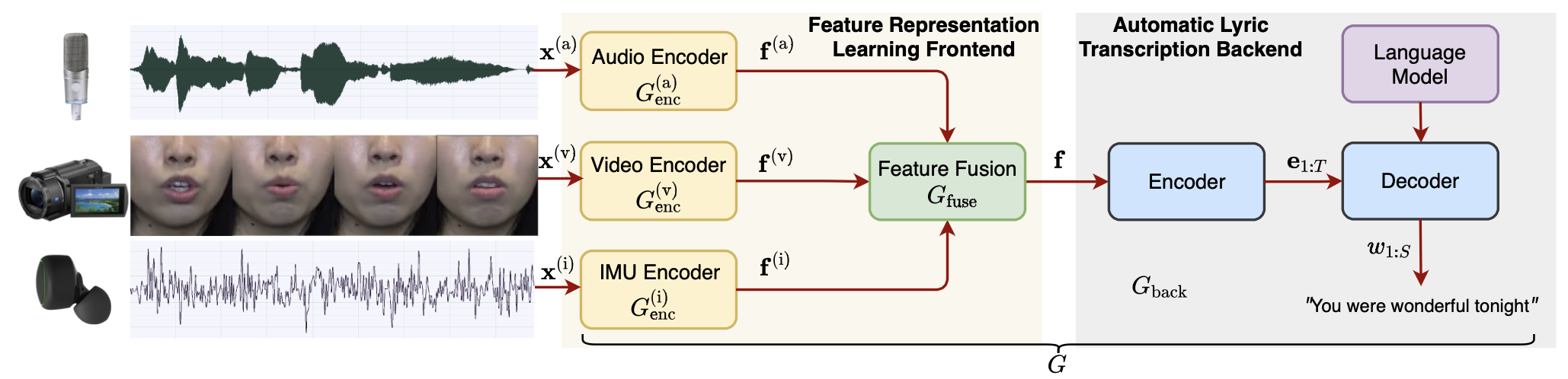}
\caption{An overview of the MM-ALT system.}
\label{fig1}
\end{center}
\vskip -0.1in
\end{figure*}

\subsection{Problem Formulation}
We consider the multimodal setting for automatic lyric transcription. More specifically, given the synchronized singing audio signal $\textbf{x}^{(\text{a})}$, video signal $\textbf{x}^{(\text{v})}$, and IMU signal $\textbf{x}^{(\text{i})}$, our goal is to transcribe the lyrics, i.e., obtain a word sequence $w_{1:S}$ representing the corresponding lyrics of the signal. As shown in Fig. \ref{fig1}, we propose the MM-ALT system to solve this problem. The system $G$ consists of a feature representation learning frontend and automatic lyric transcription backend. Firstly, modality-specific encoders $G_{\text{enc}}^{(\text{a})}, G_{\text{enc}}^{(\text{v})}, G_{\text{enc}}^{(\text{i})}$ are adopted to extract the features for each modality of signal. Then the feature fusion module $G_{\text{fuse}}$ projects the features from different modalities into the same latent space and integrates them into more representative features.  Finally, the hybrid CTC-Attention backend $G_{\text{back}}$ transforms the sequence of fused features into the lyrics. The whole system can be represented as:
\begin{align}
    w_{1:S}&=G(\textbf{x}^{(\text{a})}, \textbf{x}^{(\text{v})}, \textbf{x}^{(\text{i})})\nonumber\\
    &=G_{\text{back}}(G_{\text{fuse}}(G_{\text{enc}}^{(\text{a})}(\textbf{x}^{(\text{a})}), G_{\text{enc}}^{(\text{v})}(\textbf{x}^{(\text{v})}), G_{\text{enc}}^{(\text{i})}(\textbf{x}^{(\text{i})})))
\end{align}

\subsection{Audio Encoder}
The audio encoder $G_{\text{enc}}^{(\text{a})}$ aims to learn acoustic representations from the audio signal. Traditionally, the TDNN network and its variants dominate the field of ALT \cite{dabike2019alt:dsing,demirel2020AutomaticLyricsTranscription,ahlback2021alt:mstre}. In this paper, we propose to utilize wav2vec 2.0 \cite{baevski2020wav2vec} as the audio encoder for ALT through a transfer learning paradigm because wav2vec 2.0 generalizes well into new domains with low-resource labeled data. From our experiments, we notice that when accepting only the audio signal, our system achieves state-of-the-art performance, exceeding the results of all published approaches on the DSing dataset \cite{dabike2019alt:dsing}, one of the mainstream lyric transcription datasets.

Wav2vec 2.0 consists of a feature encoder and a context network. The feature encoder has seven blocks, each containing a temporal convolution followed by a layer normalization and a GELU activation. It takes raw audio and outputs latent speech representations. The context network contains 24 transformer blocks with model dimension 1,024, inner dimension 4,096, and 16 attention heads. It transforms latent speech representations into context representations by capturing global temporal information. Each frame of final output $\textbf{f}^{(\text{a})}$ is about 20 ms and has 1,024 dimensions. We refer readers to Appendix B for the detailed implementations. To transfer the knowledge of wav2vec 2.0 from the speech domain into the singing domain, we remove the quantization module in the original wav2vec 2.0 structure and fine-tune the model on singing datasets.  

\subsection{Video Encoder}
The video encoder $G_{\text{enc}}^{(\text{v})}$ seeks to learn visual representations of speech from cropped lip videos. Since this is the first attempt to transcribe lyrics from video modality, there are currently no benchmark models. We propose adopting the Audio-Visual Hidden Unit BERT (AV-HuBERT) \cite{shi2022avsr:avhubert} in our system, which is the state-of-the-art approach in the task of lip reading for speech recognition. 

AV-HuBERT consists of an image encoder and a backbone transformer encoder. The image encoder is built by a 3D convolutional layer followed by a ResNet-18 block \cite{he2016deep}. Similar to the context network in wav2vec 2.0, the transformer encoder has 24 blocks, each of which has a model dimension of 1,024, a feed-forward dimension of 4,096, and 16 attention heads. Each frame of resulting representations $\textbf{f}^{(\text{v})}$ is about 40 ms and has 1,024 dimensions. We remove the audio layer in the original structure and only feed the video modality into AV-HuBERT. 

\subsection{IMU Encoder}
\begin{figure}[t]
\vskip -0.2in
\begin{center}
\includegraphics[width=\linewidth]{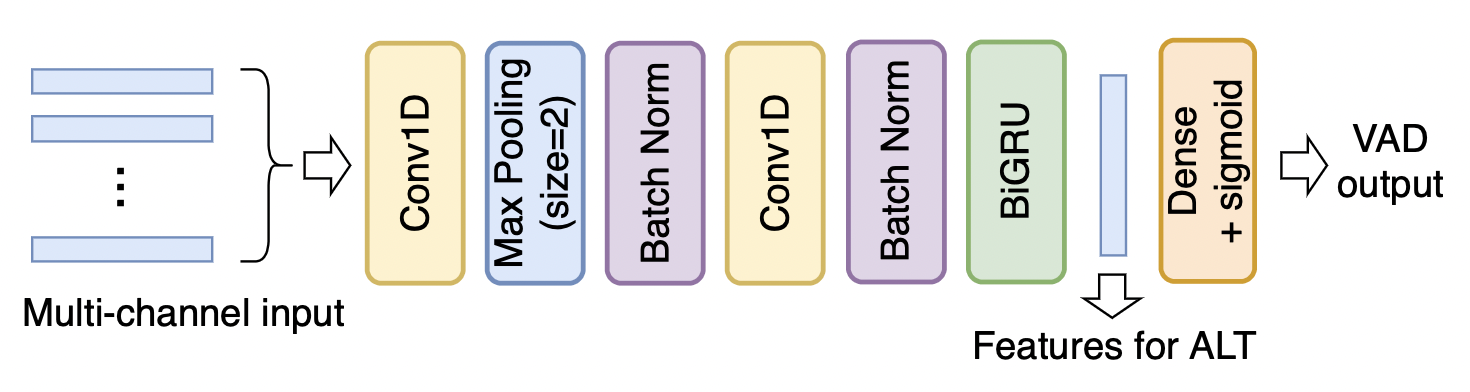}
\caption{Structure of IMU CRNN.}
\label{fig:crnn}
\end{center}
\vskip -0.2in
\end{figure}

To validate the assumption that IMU motion data correlates with the audio signal, and further exploit such correlation to help the lyric transcription, the IMU encoder $G_{\text{enc}}^{(\text{i})}$ converts IMU signal to features that correspond to speaking. We utilize a convolutional-recurrent neural network (CRNN) containing 1D convolutional layers and bi-directional Gated Recurrent Unit (GRU) layers, resembling the best-performing network CNN-BiLSTM in \cite{ott2022imu:crnn}. Fig. \ref{fig:crnn} shows the structure of our IMU CRNN. The input of CRNN contains $8$ channels, including three axes of accelerometer data, three axes of gyroscope data, and the quadratic sums of different channels of the two sensors, respectively. When incorporated into the ALT system, the output of the last GRU layer $\textbf{f}^{(\text{i})}$ is used as the IMU features. Each $\textbf{f}^{(\text{i})}$ has 20 ms frame length and 120 dimensions.

\subsection{Feature Fusion Module}
\begin{figure}[b]
\vskip -0.2in
\begin{center}
\includegraphics[width=\linewidth]{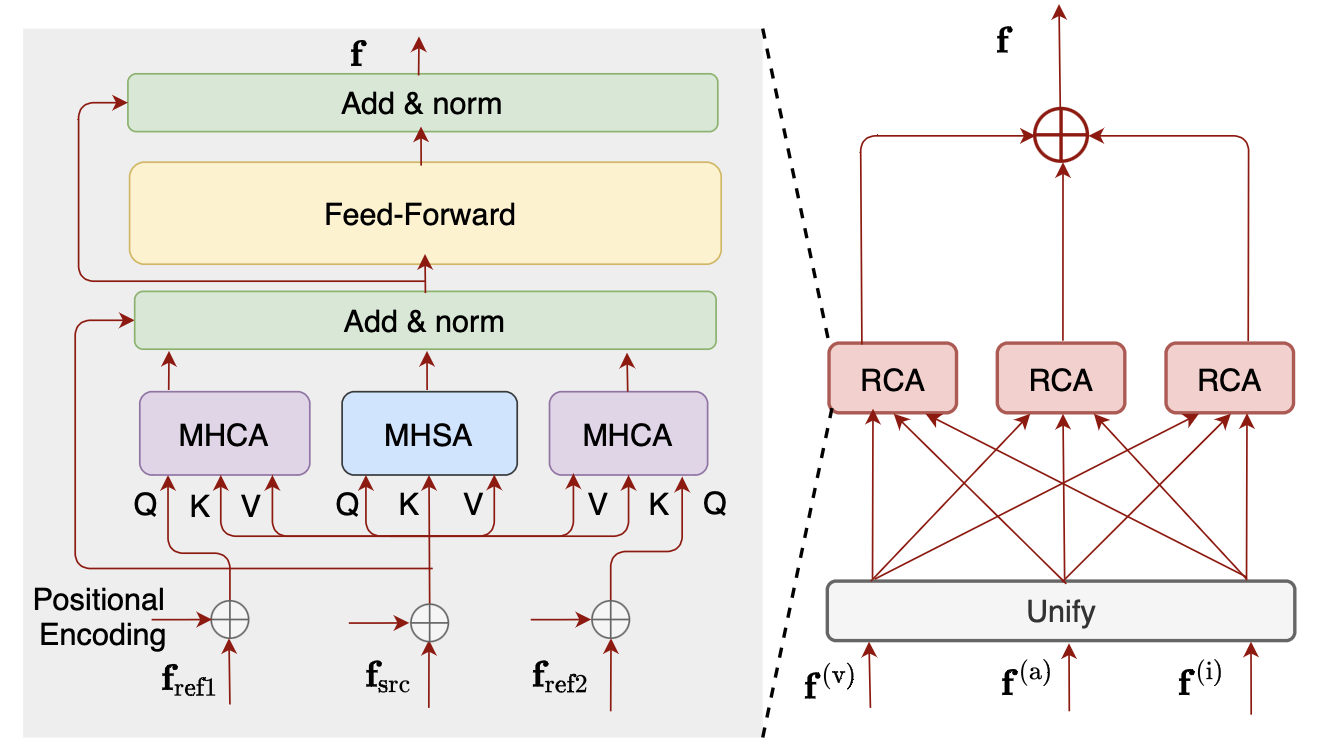}
\caption{An illustration of the feature fusion module and the proposed Residual Cross Attention (RCA) algorithm.}
\label{fig:rca}
\end{center}
\vskip -0.2in
\end{figure}

The feature fusion module $G_{\text{fuse}}$ is designed to exploit the complementary relationship and redundancy of different modalities. As illustrated in Fig. \ref{fig:rca}, the dimensions of features from different modalities are firstly unified. We up-sample the video to the same time resolution as the features of audio and IMU. Then we pad or truncate the features of the other two modalities to standardize the number of frames amongst the three modalities before the fusion operation. For IMU features, we adopt a linear layer following a GELU activation to increase its dimensions from 120 to 1024.
    
We propose a new attention module named Residual Cross Attention (RCA). RCA is built upon Transformer block architecture. An RCA block accepts inputs from multiple modalities. One input is regarded as the source, providing keys and values, while other inputs are considered as the reference, which provides queries. Besides the multi-head self-attention (MHSA) \cite{vaswani2017attention} operation on the source, RCA adds extra shortcuts by computing the multi-head cross-attention (MHCA) operation between the source and each reference. RCA can be represented by Eq. \ref{eq2} and Eq. \ref{eq3}:

    \begin{equation}
    \label{eq2}
    \begin{split}
         \textbf{f}^{\prime} =   \text{  LN}(\textbf{f}_{\text{src}}&+\text{MHSA}(\textbf{f}_{\text{src}}) 
          + \text{MHCA}(\textbf{f}_{\text{src}}, \textbf{f}_{\text{ref1}}) \\ &
          + \text{MHCA}(\textbf{f}_{\text{src}}, \textbf{f}_{\text{ref2}}))
    \end{split}
    \end{equation}

    \begin{equation}\label{eq3}
        \textbf{f}=\text{LN}(\text{FFN}(\textbf{f}^{\prime}) + \textbf{f}^{\prime})
    \end{equation}
where LN refers to a layer normalization layer, and FFN is a positional-wise feed forward network the same as the one in a Transformer block \cite{vaswani2017attention}. There are three RCA modules during the feature fusion, and each modality's input serves as the source in one of the modules. Finally, we add the output together and obtain the final fused features $\textbf{f}$.

The motivations behind the proposal of RCA in multimodal scenarios are explained as follows. Firstly, RCA adopts self-attention and residual shortcuts to extract global relationships among features of all frames and reserves information of source modality. Secondly, RCA takes advantage of complementary information from reference modalities through its cross-attention mechanism. Specifically, the queries from reference modalities help downstream modules delve into missing relationships between time frames that are not attended to when using only self-attention.

\subsection{Hybrid CTC-Attention Backend}
Inspired by \cite{watanabe2017hybrid}, we design a hybrid CTC-Attention backend to map a sequence of fused features $\textbf{f}_1,\textbf{f}_2, ...,\textbf{f}_T\in \mathbb{R}^{1024}$ into a sequence of tokens $w_1, w_2, ...,w_{S}\in\mathbb{V}$, where $\mathbb{V}$ is the vocabulary. In our implementations, $\mathbb{V}$ has 31 tokens including 30 character targets and a word boundary token. 

Firstly, we adopt a 2-layer MLP as the encoder to further encode the fused representations $\textbf{f}_{1:T}$ into $\textbf{e}_{1:T}\in \mathbb{R}^{T\times1024}$. Then we design two branch networks, one of which is used to output CTC predictions and another is used as the decoder to output Sequence-to-Sequence (S2S) predictions. Specifically, a projection layer is used to map $\textbf{e}_{1:T}$ into the output probabilities for each frame $p(\pi_t|\textbf{e}_t)$, where the CTC predictions $\pi_t\in \mathbb{V}$. $\textbf{e}_{1:T}$ are also fed into the S2S decoder, which is parameterized via a location-aware attention-based GRU \cite{chorowski2015attention}, to be autoregressively decoded into target probabilities $p(w_c|w_{1:c-1}, \textbf{e}_{1:T})$. $p(\pi_t|\textbf{e}_t)$ and $p(w_c|w_{1:c-1}, \textbf{e}_{1:T})$ are used for two different training and inference schemes, which will be elaborated in the next section.


\subsection{Training and Inference}
The whole system is trained with cross-entropy loss of outputs from both connectionist temporal classification (CTC) \cite{graves2006e2e:ctc} and sequence-to-sequence (S2S) \cite{bahdanau2016end} attention-based decoder. Suppose the ground truth transcription is $w_1, w_2,...,w_S$. The CTC loss can then be written as:
\begin{align}\label{eq4}
    \mathcal{L}_{\text{CTC}} &= -\log p_{\text{CTC}}(w_{1:S}|\textbf{e}_{1:T}) \nonumber \\
    &= -\log\sum_{\pi_{1:T}\in\mathcal{B}^{-1}(w_{1:S})}\prod_{t=1}^Tp(\pi_t|\textbf{e}_t)
\end{align}

Here, $\mathcal{B}$ is an operation mapping an alignment sequence $\pi_{1:T}$ to $w_{1:S}$ by removing repeated tokens and word boundary tokens. $\mathcal{B}^{-1}(w_{1:S})$ refers to all the CTC paths mapped from $w_{1:S}$. 

The S2S loss can be written as:
\begin{align}\label{eq5}
    \mathcal{L}_{\text{S2S}} &= -\log p_{\text{S2S}}(w_{1:S}|\textbf{e}_{1:T}) \nonumber \\
    &= -\log \prod_{c=1}^S p(w_c|w_{1:c-1}, \textbf{e}_{1:T})
\end{align}

The overall loss function is the addition of the two loss terms above. We introduce a hyper-parameter, $\alpha$, to balance both losses. 
\begin{equation}\label{eq6}
    \mathcal{L}=(1-\alpha)\mathcal{L}_{\text{S2S}} + \alpha \mathcal{L}_{\text{CTC}}
\end{equation}

During inference, the most likely lyrics will be predicted considering the output of CTC, S2S, and language model (LM), by the following equation:
\begin{align}\label{eq7}
    w_{1:S}^*=&\arg\max_{w_{1:S}}\beta\log p_{\text{CTC}}(w_{1:S}|\textbf{e}_{1:T}) \nonumber \\
    +&(1-\beta)\log p_{\text{S2S}}(w_{1:S}|\textbf{e}_{1:T}) + \gamma \log p_{\text{LM}}(w_{1:S})
\end{align}

Here, $\beta$ and $\gamma$ are two hyper-parameters used to balance three log-probability terms during the beam search. When $\gamma > 0$, the LSTM language model is enabled. Beam size is set as 512.


\section{Experiments}

In this section, we evaluate our proposed MM-ALT system using curated N20EM dataset. Specifically, we firstly conduct single-modality experiments to evaluate the modality-specific representation learning. Then we evaluate the whole system in multimodal scenarios to demonstrate its effectiveness. Finally, we simulate the realistic environments by adding musical accompaniments as perturbations to test the robustness of our MM-ALT system. 

\subsection{Implementation Details}
We build our MM-ALT system using the PyTorch library and SpeechBrain toolkit \cite{ravanelli2021speechbrain}. As for more detailed model configurations, please refer to Appendix B. We apply data augmentation during training: for the audio signal, we perform SpecAugment \cite{park2019specaugment} in the time domain; for the video signal, we randomly flip and crop face images with the size of 88 following \cite{shi2022avsr:avhubert}. All models are trained using the Adam optimizer. For the ALT task, we report the Word Error Rate (WER) as evaluation metrics.

\subsection{Single-Modality Tasks}
\label{sec:single}

\subsubsection{Automatic Lyric Transcription} 
We evaluate the performance of audio encoder together with the automatic lyric transcription backend on the curated N20EM dataset and DSing \cite{dabike2019alt:dsing} dataset, one of the mainstream ALT datasets. Firstly, wav2vec 2.0 \cite{baevski2020wav2vec} is pretrained on LibriVox (LV-60k) and loaded into our audio encoder\footnote{\url{https://huggingface.co/facebook/wav2vec2-large-960h-lv60-self}}. Then we train the models on training split, validate/test the ALT performance on validation/test split. In inference, we use CTC-S2S-LM to decode the lyrics. For the DSing dataset, we train an LSTM LM only on text corpus from DSing training split, which is a subset of text corpus used by other baselines. While for the N20EM dataset, we train an LSTM LM on text corpus from LibriSpeech \cite{panayotov2015librispeech}, DSing \cite{dabike2019alt:dsing} and the N20EM dataset. We keep this LM configurations for other experiments on the N20EM dataset. Besides, we enable the training of the fusion module and set $\textbf{f}^{(v)}$ and $\textbf{f}^{(i)}$ as zero tensors for a fair comparison with the multimodal settings. 

The evaluation results are summarized in Table \ref{tbl-audio}. We observe that our proposed system achieves 13.26\% and 14.56\% WER on the DSing dataset, outperforming TDNN and its variants \cite{dabike2019alt:dsing,demirel2020AutomaticLyricsTranscription,ahlback2021alt:mstre} by at least 4.44\% on the validation split and 0.40\% on the test split. The performance of our system indicates that we can successfully adapt the pre-trained model from the speech domain to the singing domain by exploiting the similarities between speech and singing. 

\begin{table}[t]
\caption{\textbf{WER(\%) of ALT systems on Singing datasets.} We compare different methods on DSing dataset and build benchmarks for N20EM dataset. ``w. DSing'' refers to adding DSing to the training data.}
	\centering
	\begin{tabular}{l|l|c|c}
		\toprule
		Method & Dataset & Validation & Test \\
		\midrule
		TDNN-F \cite{dabike2019alt:dsing} & DSing & 23.33 & 19.60 \\
		CTDNN-SA \cite{demirel2020AutomaticLyricsTranscription} & DSing & 17.70 & 14.96 \\
		MSTRE-Net \cite{ahlback2021alt:mstre} & DSing & - & 15.38 \\
		Ours & DSing & \textbf{13.26} & \textbf{14.56} \\
		\midrule
		Ours & N20EM & 12.74 & 19.68 \\
		Ours w. DSing & N20EM & \textbf{9.65} & \textbf{13.00}\\
		\bottomrule
	\end{tabular}
    
    \label{tbl-audio}
\end{table}


Since our proposed ALT system displays great capability on the DSing dataset, we adopt it as a strong baseline for the N20EM dataset. To begin with, we train the system using only the N20EM dataset and observe $12.74\%$ WER on the validation split and $19.68\%$ WER on the test split. When adding the DSing dataset to the training data, performance is improved to $9.65\%$ and $13.00\%$ WER on validation and test splits respectively, demonstrating that having more singing data during the training enhances the ability of the system to generalize.

\subsubsection{Lyric Lipreading} We initialize a new task in this section, named as lyric lipreading. This task aims to recognize lyrics only through video modality. Since this is the very first attempt, we train our video encoder together with automatic lyric transcription backend to build a benchmark system. Firstly, AV-HuBERT \cite{shi2022avsr:avhubert} is pretrained on LRS3 \cite{afouras2018lrs3} and VoxCeleb2 \cite{chung2018voxceleb2}, and then loaded into our video encoder\footnote{\url{https://dl.fbaipublicfiles.com/avhubert/model/lrs3_vox/vsr/self_large_vox_433h.pt}}. Afterwards, we train the models on the N20EM dataset. Unlike the original implementation of unigram tokenizer in \cite{shi2022avsr:avhubert}, we adopt a character tokenizer as mentioned in Section 4.6 to facilitate feature fusion and comparison among different modalities. 

\begin{table}[t]
\caption{\textbf{WER(\%) of lyric lipreading on N20EM dataset.} We compare different decoding configurations.}
	\centering
	\begin{tabular}{c|c|c|c|c}
		\toprule
		CTC &  S2S & LM & Validation & Test \\
		\midrule
		$\surd$ & $\times$ & $\times$ &  63.52 & 78.20 \\
		$\times$ & $\surd$ & $\times$ &  55.72 & 74.10 \\
		$\surd$ & $\surd$ & $\times$ &  55.80 & 72.70 \\
		$\surd$ & $\surd$ & $\surd$ &  \textbf{47.91} & \textbf{68.45} \\
		\bottomrule
	\end{tabular}
    
    \label{tbl-video}
\end{table}

During experiments, we evaluate different decoding configurations. As shown in Table \ref{tbl-video}, we observe that the performance of the S2S model exceeds the CTC model by a large margin. Furthermore, adopting a trained language model can also enhance system performance. Our best model achieves $47.91\%$ WER on the validation split and $68.45\%$ WER on the test split. 

\subsubsection{Voice Activity Detection from IMU}

From our experiments, it is difficult to obtain a functional IMU feature extractor by directly training the IMU encoder on the ALT task, because IMU signal is not as informative as acoustic data when the goal is to predict words. To maximize the use of IMU data, rather than immediately tackling the problem of ALT with only IMU data, we first attempt a relevant task of lower complexity: voice activity detection (VAD), the task of detecting zones of vocal activity. VAD with audio as inputs is usually an essential step to preprocess data for ASR systems to lower the downstream processing latency \cite{chang2018vad:google, ariav2019vad:multimodal}. Inspired by this, we hope to provide additional meaningful inductive bias for consequent transcription modules by performing VAD training on the IMU encoder. We integrate the IMU encoder into our multimodal ALT system after it is well-trained on the VAD task.


The annotation of VAD is obtained from audio recordings. For each recording, we select a loudness threshold that is slightly higher than the loudness of the sound of breaths. We get a rough VAD annotation by treating all frames with a volume above this threshold as voiced and the rest as unvoiced. Afterward, IMU signals of songs are segmented into 5-second clips for training.

\begin{table}[t]
\caption{\textbf{Results of frame-level classification for VAD using IMU signal.}}
	\centering
	\begin{tabular}{l|c|c}
		\toprule
		Metric (\%) & Validation & Test \\
		\midrule
		Accuracy & 69.63 & 74.36 \\
		F1-score & 79.72 & 82.86\\
		Macro F1-score & 62.71 & 67.09 \\
		\bottomrule
	\end{tabular}
    
    \label{tb_imu}
\end{table}

\begin{figure}[t]
\vskip -0.2in
\begin{center}
\includegraphics[width=\linewidth]{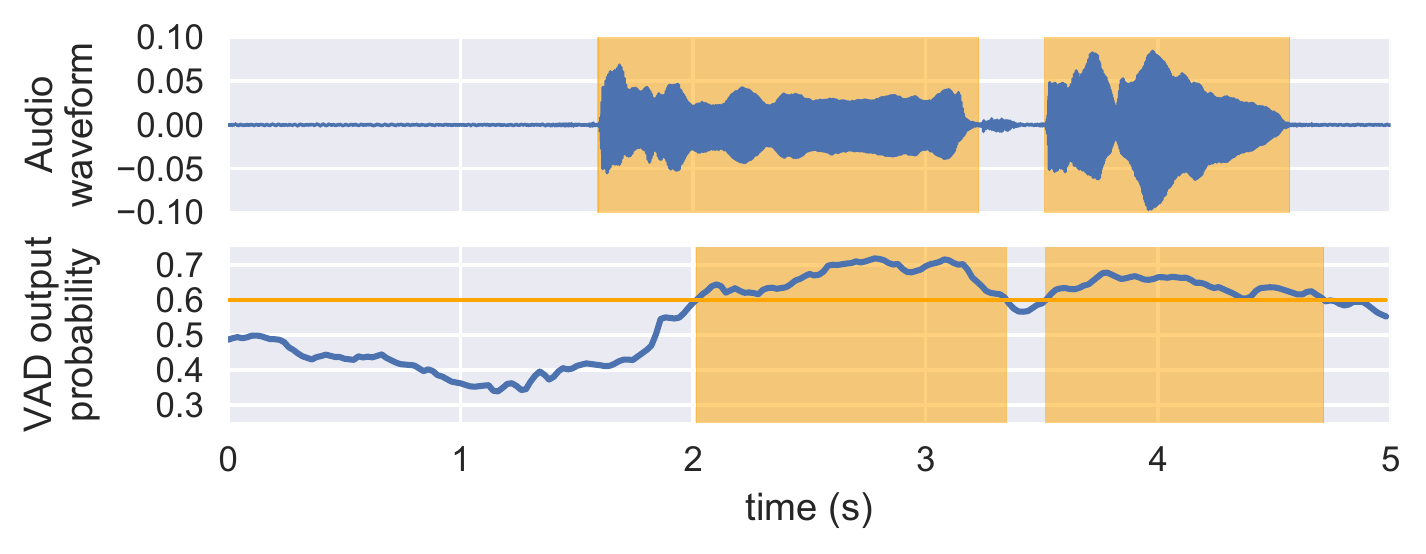}
\caption{An audio clip and the corresponding IMU VAD's output. Highlighted area represents actual/predicted voiced segments. The orange horizontal line in the VAD output figure is the classification threshold.}
\label{fig:crnn-exp}
\end{center}
\vskip -0.2in
\end{figure}

Table \ref{tb_imu} shows the results of our IMU VAD model. On the test split, we achieve an accuracy of 74.36\%, an F1 score of 82.86\%, and a 67.09\% macro F1. These results suggest the correlation between the wearable IMU sensor and microphone recording data, and our model can successfully capture this correlation. Fig. \ref{fig:crnn-exp} shows the relationship between audio waveform and IMU VAD output in a more intuitive way. We can find that VAD output probabilities are low at unvoiced frames while becoming higher when audio frames contain voice activities.

\begin{table}[b]
\caption{\textbf{WER(\%) of various modality combinations in different SNR scenarios on N20EM dataset.} A: Audio-only ALT, A-V: Audio-Visual ALT, A-V-I: Audio-Visual-IMU ALT.}
	\label{tbl-multimodal}
	\centering
	\begin{tabular}{c|ccc|ccc}
		\toprule
		SNR (dB) & \multicolumn{3}{c|}{Validation} & \multicolumn{3}{c}{Test} \\
        & A & A-V & A-V-I & A & A-V & A-V-I\\
		\midrule
		-10 & 74.33 & 42.50 & \textbf{41.87} & 90.96 & 64.83 & \textbf{63.92} \\
		-5 & 45.45 & 36.91 & \textbf{36.91} & 63.02 & 54.68 & \textbf{54.68} \\
		0 & 25.28 & \textbf{22.70} & 22.81 & 35.70 & \textbf{34.42} & 34.44\\
		5 & 16.06 & 15.57 & \textbf{15.06} & \textbf{22.19} & 22.38 & 22.89\\
		10 & \textbf{12.37} & 12.96 & 12.51 & 17.41 & 17.41 & \textbf{17.03}\\
		$\infty$ (clean) & 9.65 & 9.56 & \textbf{9.45} & 13.00 & 12.81 & \textbf{12.71} \\
		\midrule
		Avg. & 30.52 & 23.37 & \textbf{23.10} & 40.38 & 34.42 & \textbf{34.27} \\
		\bottomrule
	\end{tabular}
    
\end{table}
\subsection{Multi-Modality ALT}
We evaluate our multimodal ALT system in Fig. \ref{fig1} on the N20EM dataset. Before training the whole system, we adopt a transfer learning strategy by loading the modality-specific encoders trained in single-modality experiments. Unlike in section \ref{sec:single}, we also simulate realistic environments by adding corresponding musical accompaniments to the audio signal with different signal-to-noise ratios (SNRs). We train and evaluate our system on the mixed data above to test system robustness. We compare different modality combinations in $-10, -5, 0, 5, 10\ \text{dB}$ SNRs, as well as clean scenarios (no accompaniments). 

\subsubsection{Quantitative Analysis}
Quantitative results are reported in Table \ref{tbl-multimodal}. Firstly, we compare the audio-only system with the audio-visual system. We notice that the audio-visual system performs much better than audio-only in low SNR scenarios. For example, in $-10\ \text{dB}$, the audio-visual system significantly outperforms the audio-only system by about $30\%$ WER. As the SNR level increases, the performance of the two systems gradually converges. When there are no accompaniments, an audio-visual system achieves $9.56\%$ WER on the validation split and $12.81\%$ WER on the test split, which exceeds the performance of an audio-only system. We also report the average WER on the above six scenarios and find that the audio-visual system outperforms the audio-only system by $7.15\%$ and $5.96\%$ on the validation and test splits respectively, demonstrating its superior robustness.

After adding the IMU modality, we notice that our 3-modal MM-ALT system achieves the best results in the majority of scenarios. The MM-ALT system achieves $23.10\%$ and $34.27\%$ WER on the validation and the test splits on average, surpassing both audio-only and audio-visual configurations. Without accompaniments, our multimodal system achieves $9.45\%$ WER on the validation split and $12.71\%$ WER on the test split, which is the best result obtained for the N20EM dataset. Admittedly, the improvements brought by IMU modality is not significant, which indicates that IMU modality is not as promising as video modality. However, its effectiveness is still not negligible. We leave it for future work to maximize the effectiveness of the IMU modality.

\begin{table}[t]
\caption{Qualitative Results. Insertions are marked in \textcolor{red}{\underline{underline and red}}, and substitutions are marked in \textcolor{cyan}{\textit{italics and cyan}}.}
\label{tb:case}
\vskip -0.1 in
\begin{tabular}{@{}l|c|l@{}}
\toprule
                           &                                & \multicolumn{1}{c}{Text}       \\ \midrule
\multicolumn{1}{c|}{} & \multicolumn{1}{c|}{Reference} & Son of god love’s pure light   \\
\multicolumn{1}{l|}{\multirow{2}{*}{Clean}}      & \multicolumn{1}{c|}{A}         & \textcolor{red}{\underline{The}} son of god \textcolor{cyan}{\textit{loves}} pure \textcolor{cyan}{\textit{life}} \\
\multicolumn{1}{l|}{}      & \multicolumn{1}{c|}{A-V}       & Son of god \textcolor{cyan}{\textit{loves}} pure \textcolor{cyan}{\textit{life}}     \\
\multicolumn{1}{l|}{}      & \multicolumn{1}{c|}{A-V-I}     & Son of god \textcolor{cyan}{\textit{loves}} pure \textcolor{cyan}{\textit{ligh}}    \\ \midrule
\multicolumn{1}{c|}{} & \multicolumn{1}{c|}{Reference} & Wonder how I got along         \\
\multicolumn{1}{l|}{\multirow{2}{*}{Noisy}}      & \multicolumn{1}{c|}{A}         & \textcolor{cyan}{\textit{What}} \textcolor{red}{\underline{is}} how I got \textcolor{cyan}{\textit{a}} \textcolor{red}{\underline{lot}}        \\
\multicolumn{1}{l|}{}      & \multicolumn{1}{c|}{A-V}       & \textcolor{cyan}{\textit{Wander}} how I got \textcolor{cyan}{\textit{a}} \textcolor{red}{\underline{long}}        \\
\multicolumn{1}{l|}{}      & \multicolumn{1}{c|}{A-V-I}     & Wonder how I got \textcolor{cyan}{\textit{a}} \textcolor{red}{\underline{long}}        \\ \bottomrule
\end{tabular}
\vskip -0.1 in
\end{table}

\subsubsection{Qualitative Analysis} 
Apart from the quantitative analysis, we also show the qualitative results in Table \ref{tb:case}. We show one case from the clean scenario and one from a noisy (mixed with accompaniments) scenario. More quantitative results are displayed in Appendix C. In the clean case, the audio-only system has three word errors, including one insertion and two substitutions. Both audio-visual and audio-visual-IMU models misspell ``love's'' and ``light'', but for the word ``light'' the three-modal system's output has fewer character-level errors (only one missing character)
. Likewise, in the noisy case, the MM-ALT system also performs better than its audio-only and audio-visual counterparts by correcting the substitutions of "What" and "Wander" as well as the insertion of "is". Although the word "along" in reference is not fully recovered, the transcription from the MM-ALT system is closer to the ground truth than the transcription from the audio-only system.

\subsubsection{Ablation Study}
To validate the effectiveness of our proposed RCA mechanism, we conduct an ablation study for the feature fusion module in our MM-ALT system. The results are summarized in Table \ref{tbl_ablation}. To magnify the differences, we evaluate the ALT performance in $-10\ \text{dB}$ SNR scenario. We find that without cross attention shortcuts, the ALT performance will drop $0.6\%$ and $3.01\%$ WER on the validation and test splits. Without a self-attention mechanism, the ALT performance will decrease by $0.8\%$ and $0.55\%$ WER respectively. These results suggest that RCA contributes to better feature fusion.

\begin{figure}[t]
\vskip -0.3in
\begin{center}
\includegraphics[width=\linewidth]{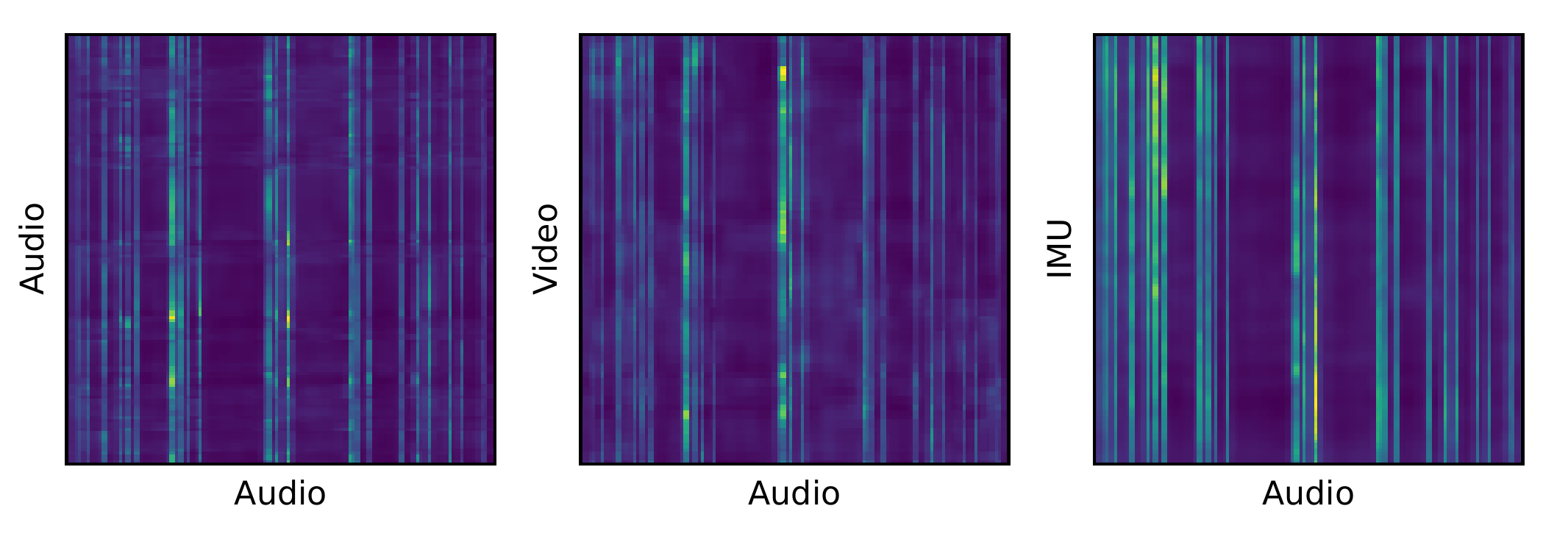}
\caption{Visualization of self-attention and cross-attention weights in RCA module of audio modality. We use brighter colors to highlight stronger attention.}
\label{fig:attnmap}
\end{center}
\vskip -0.3 in
\end{figure}

To further prove the effectiveness of the RCA mechanism, we visualize the attention maps in the RCA module when audio modality serves as the source, including self-attention and cross attention maps, as shown in Fig. \ref{fig:attnmap}. We observe that these three attention maps have captured common attention patterns. Furthermore, audio-video cross attention and audio-IMU cross attention can also extract missing relationships between time frames that are not captured by self-attention.

\begin{table}[t]
\caption{Ablation study of the Residual Cross Attention (RCA) in -10dB scenario of three modalities. CA: Cross Attention, SA: Self Attention.}
	\centering
	\vskip -0.1in
	\begin{tabular}{c|c|c}
		\toprule
		Method & Validation & Test \\
		\midrule
		w/o CA & 42.47 & 66.93 \\
		w/o SA & 42.67 & 64.47\\
		\textbf{RCA} & \textbf{41.87} & \textbf{63.92} \\
		\bottomrule
	\end{tabular}
    \vskip -0.1in
    \label{tbl_ablation}
\end{table}

\section{Conclusion} 

In this paper, we proposed a multimodal lyric transcription system, MM-ALT. Firstly, we curated a multimodal ALT dataset N20EM to facilitate building the system. Then, we presented an audio-based transcription method that surpasses existing state-of-the-art approaches. Next, we built modality-specific encoders based on video and IMU modalities. For video, we attempted the new task of lyric lipreading. For the IMU modality, we made the first attempt at frame-level VAD, which showed promising results and demonstrated the correlation between IMU and audio data. Finally, we used RCA, our proposed feature fusion method, to fuse information from different modalities to obtain the final transcription. Our series of attempts revealed that introducing additional modalities improves transcription performance and makes the model more robust to perturbations of musical instruments.



\begin{acks}
We would like to thank anonymous reviewers for their valuable suggestions. We also appreciated Sng Jia Ming Fadi Faris's work for our project website. This project is funded in part by a grant (R-252-000-A56-114) from Singapore Ministry of Education.
\end{acks}

\bibliographystyle{ACM-Reference-Format}
\bibliography{_main}

\appendix

\section{Annotate pronunciation errors in dataset}

The paradigm used for resolving linguistic errors in utterances is illustrated in Table 1. We use different brackets to represent different error types. At the phoneme level, mispronounced words were marked with front slashes when they were found to be poorly enunciated or simply wrong. Simple substitutions were noted by including the original word in square brackets following its substitute. Inserted words were marked with curly brackets in their utterances, and deleted words were put in parentheses. 

\begin{table}[htb]
\caption{Annotation for differnet types of pronunciation error.}
\centering
\vskip -0.1 in
 \begin{tabular}{ l | c | c } 
  \toprule
  Error & Example & Resolution \\ 
  \midrule
  Mispronunciation & "little" - "laytle" & /little/ \\ 

  Substitution & "the" - "a" & a [the] \\ 

  Insertion & "" - "a" & \{a\} \\ 

  Deletion & "and" - "" & (and) \\ 
  \bottomrule
 \end{tabular}
 \vskip -0.1 in
 \label{table:1}
\end{table}

\section{Implementation details}

\subsection{wav2vec 2.0}
We have included a brief description of the wav2vec 2.0 structure in Section 4.2 Audio Encoder. It consists of a feature encoder and a context network \cite{baevski2020wav2vec}. The feature encoder is a multi-layer convolutional neural network consisting of seven blocks. Each block has a 1D temporal convolution followed by a layer normalization \cite{ba2016layer} and a GELU activation \cite{hendrycks2016gaussian}. The 1D temporal convolutions have 512 channels, \{10,3,3,3,3,2,2\} kernel sizes and \{5,2,2,2,2,2,2\} strides for each block. Suppose the input to the feature encoder is $\textbf{x}$, which is 1D raw audio; its output $\textbf{z}$ is a 2D latent representation of the input $\textbf{x}$. The frequency of $\textbf{z}$ is 49 Hz, and each frame has 1,024 dimensions. After that, $\textbf{z}$ are fed into the context network to obtain the final output $\textbf{f}$, which has been explained in Section 4.2. In addition to these two parts, wav2vec 2.0 also has a quantization module for unsupervised pretraining. Through this quantization module, latent speech representation $\textbf{z}$ are discretized to $\textbf{q}$, which serve as the targets for unsupervised pretraining. Since we don't need the pretraining stage, we remove the quantization module in the original structure in our experiments.

\subsection{AV-HuBERT}
We have included a concise description of the AV-HuBERT structure in Section 4.3 Video Encoder. The original structure is able to accept both audio and video modalities \cite{shi2022avsr:avhubert}. Since we only need AV-HuBERT to extract the visual representations, we remove layers handling audio inputs. The video clips are firstly fed into a 3D convolutional frontend. The 3D convolutional frontend has a 3D convolution with $5\times7\times7$ kernel size and $1\times2\times2$ stride, a batch normalization \cite{ioffe2015batch}, a ReLU activation and a 3D max pooling  with $1\times3\times3$ kernel size and $1\times2\times2$ stride. Then a modified ResNet-18 \cite{he2016deep} is adopted to extract the latent visual features. To compensate for the absence of audio modality, we concatenate the latent visual features with zero tensors and remove the modality dropout in the original structure. The fused features are fed into the context network to obtain the final output, which has been elaborated in Section 4.3. 

\subsection{IMU CRNN}
In IMU CRNN (see section 4.4 IMU Encoder), the 1D convolutions have \{128, 200\} channels for each layer, kernel size of 3, and stride of 1. The strides for max-pooling layers are 2. ReLU activation is applied after each layer. Dropout \cite{srivastava2014dropout} is applied after each convolution and GRU layer with rates of 0.5 and 0.2, respectively. The GRU contains two layers with 60 hidden units in each layer. 

\subsection{Experiments}
For the Voice Activity Detection (VAD) task, we train the model for $50$ epochs, evaluate it at the end of each epoch, and select the model with the lowest validation loss. The model is trained using AdamW optimizer \cite{loshchilov2017decoupled}, with $1\times10^{-3}$ learning rate.

For the Automatic Lyric Transcription (ALT) task, we train the model for $20$ epochs using Adam optimizer \cite{kingma2014adam}. Since we adopt the transfer learning paradigm for wav2vec 2.0 and AV-HuBERT, a small learning rate with $1\times10^{-5}$ is used to prevent catastrophic forgetting. For the rest parts of the ALT system, the learning rate ranges from $1\times10^{-5}$ to $5\times10^{-4}$.


\section{More Qualitative Results}

\begin{table}[tb]
\caption{More Qualitative Results. Deletions are marked in \textcolor{red}{(brackets and red)}, and substitutions are marked in \textcolor{cyan}{\textit{italics and cyan}}.}
\label{tb:more_case}
\begin{tabular}{@{}c|l|l@{}}
\toprule
SNR              & \multicolumn{1}{c|}{} & Text                                        \\ \midrule
                 & Ref.                  & Goodbye papa please pray for me             \\
\multirow{2}{*}{$\infty$ (Clean)} & A    & Goodbye \textcolor{cyan}{\textit{pop}} please pray for me              \\
                 & A-V                   & Goodbye \textcolor{cyan}{\textit{bap}} please pray for me              \\
                 & A-V-I                 & Goodbye papa please pray for me             \\ \midrule
                 & Ref.                  & Jesus Lord at thy birth                     \\
\multirow{2}{*}{10 dB}            & A    & Jesus \textcolor{cyan}{\textit{love}} \textcolor{red}{(at)} \textcolor{cyan}{\textit{the}} birth                        \\
                 & A-V                   & Jesus Lord \textcolor{red}{(at)} \textcolor{cyan}{\textit{the}} birth                        \\
                 & A-V-I                 & Jesus Lord \textcolor{red}{(at)} thy birth                        \\ \midrule
                 & Ref.                  & But the wine and the song                   \\
\multirow{2}{*}{5 dB}             & A    & \textcolor{cyan}{\textit{If}} the \textcolor{cyan}{\textit{wind}} \textcolor{cyan}{\textit{in}} \textcolor{cyan}{\textit{my}} \textcolor{cyan}{\textit{soul}}                      \\
                 & A-V                   & \textcolor{cyan}{\textit{Baby}} the \textcolor{cyan}{\textit{wind}} \textcolor{cyan}{\textit{in}} the \textcolor{cyan}{\textit{sun}}                    \\
                 & A-V-I                 & \textcolor{cyan}{\textit{Above}} the wine and the \textcolor{cyan}{\textit{soul}}                 \\ \midrule
                 & Ref.                  & Like the seasons have all gone              \\
\multirow{2}{*}{0 dB}             & A    & Like the \textcolor{cyan}{\textit{season's}} \textcolor{cyan}{\textit{hold}} \textcolor{red}{(all)} \textcolor{cyan}{\textit{on}}                   \\
                 & A-V                   & Like the seasons \textcolor{cyan}{\textit{held}} \textcolor{red}{(all)} \textcolor{cyan}{\textit{on}}                    \\
                 & A-V-I                 & Like the seasons \textcolor{cyan}{\textit{having}} \textcolor{red}{(all)} gone                \\ \midrule
                 & Ref.                  & You gave me love and helped me find the sun \\
\multirow{2}{*}{-5 dB}            & A    & You \textcolor{cyan}{\textit{give}} me love and \textcolor{cyan}{\textit{help}} me \textcolor{cyan}{\textit{open}} \textcolor{red}{(the)} \textcolor{cyan}{\textit{sew}}       \\
                 & A-V                   & You \textcolor{cyan}{\textit{give}} me love and \textcolor{cyan}{\textit{help}} me \textcolor{cyan}{\textit{in}} the \textcolor{cyan}{\textit{song}}    \\
                 & A-V-I                 & You gave me love and \textcolor{cyan}{\textit{help}} me \textcolor{cyan}{\textit{in}} the \textcolor{cyan}{\textit{song}}    \\ \midrule
                 & Ref.                  & Sleep in heavenly peace                     \\
\multirow{2}{*}{-10 dB}           & A    & \textcolor{cyan}{\textit{Edelweiss}} \textcolor{cyan}{\textit{edelweiss}} \textcolor{red}{(heavenly)} \textcolor{red}{(peace)}                         \\
                 & A-V                   & \textcolor{cyan}{\textit{Sleigh}} in \textcolor{cyan}{\textit{heaven}} \textcolor{cyan}{\textit{sleigh}}                     \\
                 & A-V-I                 & Sleep in \textcolor{cyan}{\textit{heaven}} \textcolor{cyan}{\textit{sleigh}}                      \\ \bottomrule
\end{tabular}
\end{table}

More qualitative results are displayed in Table \ref{tb:more_case}. It is noticed that our MM-ALT system performs better than its audio-only and audio-visual counterparts by reducing word errors.

\end{document}